# ET-AL: entropy-targeted active learning for bias mitigation in materials data

Hengrui Zhang 张恒睿,[1] Wei (Wayne) Chen 陈伟,[1] James M. Rondinelli,[2,*] Wei Chen[1,*]

[1]Department of Mechanical Engineering, Northwestern University, Evanston, IL 60208, USA

[2]Department of Materials Science and Engineering, Northwestern University, Evanston, IL 60208, USA

*Correspondence: jrondinelli@northwestern.edu and weichen@northwestern.edu

## ABSTRACT

Growing materials data and data-driven informatics drastically promote the discovery and design of materials. While there are significant advancements in data-driven models, the quality of data resources is less studied despite its huge impact on model performance. In this work, we focus on data bias arising from uneven coverage of materials families in existing knowledge. Observing different diversities among crystal systems in common materials databases, we propose an information entropy-based metric for measuring this bias. To mitigate the bias, we develop an entropy-targeted active learning (ET-AL) framework, which guides the acquisition of new data to improve the diversity of underrepresented crystal systems. We demonstrate the capability of ET-AL for bias mitigation and the resulting improvement in downstream machine learning models. This approach is broadly applicable to data-driven materials discovery, including autonomous data acquisition and dataset trimming to reduce bias, as well as data-driven informatics in other scientific domains.

## INTRODUCTION

Data-driven autonomous materials design has recently emerged as a new paradigm for materials discovery[1-3]. With large materials data and powerful informatics tools, this paradigm significantly accelerates the understanding of physical and chemical mechanisms in materials science[4-6], accurate prediction of materials structures and properties[7-10], as well as the design of materials with desired properties[11-14]. While the informatics tools, such as machine learning (ML) and design optimization models, hold a conspicuous position in these works, the data resources are as important[15,16]. The performances that the models can attain highly depend on the quality of data they are built upon. Data veracity entails a description of where and how data were collected, but less frequently is why (or why not) using certain data clearly articulated.

Following the Materials Genome Initiative[17], multiple materials data resources have emerged. The Materials Project[18], Open Quantum Materials Database (OQMD)[19,20], the Automatic Flow for Materials Discovery (AFLOW)[21], and the Joint Automated Repository for Various Integrated Simulations (JARVIS)[22] are prominent examples. These platforms use high-throughput first-principles calculations to evaluate various properties for a wide range of materials (stoichiometric and defect-free) and make the data publicly available. Besides these centralized data resources, a growing portion of materials data is generated in various research projects, available from published papers (including their associated repositories) and platforms such as the Materials Data Facility[23]. These distributed data are increasingly utilized owing to data/text mining tools. However, it is common that materials data do not have uniform coverage for multiple reasons: (1) The candidate materials for database construction are selected among known structures or based on known structural prototypes, and lower symmetry structures are less explored than higher symmetry ones. (2) Most literature only reports compounds perceived to exhibit "good" properties based on the aspect of interest[24], while the "unsatisfactory" results can also be valuable[25]. (3)

Property simulation is easier for compounds that are structurally simple, and property measurement is simpler for compounds that are readily synthesizable and stable at ambient pressures and temperatures. These, among other factors, lead to bias in the materials data platforms.

Data bias, a ubiquitous issue in data science, has been more recognized in the social science domain[26,27] but is often overlooked in physical sciences, including materials science. Just as it causes social inequity in social policy built upon that data, bias in materials data is harmful to data-driven materials modeling and design. Belviso et al.[28] demonstrated how bias in the chemistry space prevents an ML model from accurately predicting electronic bandgaps. Bias in the structure space is less explicit but also detrimental. An example is a bias in stability data among crystal structures, which we refer to as "structure–stability bias". Such bias hinders the modeling of phase stabilities, thus affecting the accurate prediction of microstructure. As Molkeri et al.[29] found, microstructure information is important for the modeling of various materials properties, therefore, the impact of structure–stability bias is not limited to stability itself but also on other properties.

Although some attempts have been pursued to characterize bias on trained models *post facto*[30] or reduce the impact of data bias on model training[31,32], few have addressed bias intrinsic to the data for which the models are trained and mitigated bias from the onset. The presence of bias in materials data may be inevitable since the distributions of properties are unknown and can be uneven in nature. Nonetheless, detecting the bias of datasets could alert users of their potential impact. As bias originates from uneven coverage of different materials families, it can be captured by examining the diversities of families in the data, which reflects the completeness of coverage. Moreover, by adding well-selected new data points, bias in a dataset can be reduced. Towards this end, the active learning (AL) method provides a way to sequentially select optimal data points guided by sampling strategies considering uncertainty, diversity, or performance[33-35]. AL-based methods have been applied to accelerate materials discovery targeting high performance[36-39] and chemical uniqueness[40], as well as to assess the selection of design space[41]. With a specially designed sampling strategy, AL can serve as a method for bias reduction.

In this work, we propose entropy-targeted active learning (ET-AL) as a systematic approach to detecting and reducing materials data bias. We focus on the structure–stability bias in DFT-generated databases as a use case for demonstrating the approach. With information entropy as a diversity metric, we quantify the bias of stability by its diversity among structures. We then develop an active learning method with a sampling strategy towards increasing the diversity of stability of underrepresented structures, thus reducing the bias. We demonstrate the capability of ET-AL through experiments performed on existing datasets. We show that ET-AL provides a general method for mitigating bias in materials datasets and is also applicable in guiding the construction of materials databases, thus granting materials researchers access to low-bias data for machine learning.

## RESULTS AND DISCUSSION

### Data Bias Characterization

For demonstration purposes, we retrieve two materials datasets: (1) structure and formation energy per atom of all binary intermetallic compounds among the elements Al, Ti, Cr, Fe, Co, Ni, Cu, and W from OQMD (denoted OQMD-8, size 2,953); and (2) all entries with elastic moduli available from the JARVIS classical force-field inspired descriptors (CFID) dataset[42], cleaned as described in the Methods (denoted J-CFID, size 10,898). We show the distribution of formation energy per atom $\Delta E$ of materials in the two datasets with respect to crystal system in Figure 1a–b. We consider compounds in the cubic, hexagonal, trigonal, tetragonal, and orthorhombic systems to be higher in symmetry than those of monoclinic or triclinic systems, because they possess one or more rotation axes, and their unit cells have three or fewer free interaxial angles and lattice parameters. Among the seven crystal systems, the lower symmetry monoclinic and triclinic systems display higher distribution density in the more stable (lower $\Delta E$) region. This observation contradicts the empirical rules that materials with higher symmetry (which are usually more close-packed and have higher coordination numbers) generally have higher stability[43,44]. Such contradiction is due to the imbalanced coverage of different crystal systems in the materials datasets, and we refer to this problem as "structure–stability bias".

Without assuming any prior knowledge such as the correlation between symmetry and stability, are we still able to capture the bias? To that end, we first define the diversity of a dataset by recognizing that for values of a continuous variable $Y$, the diversity can be quantified by information entropy[45]

$$h(Y) = \mathrm{E}[-\log f(y)] = -\int f(y) \log f(y) \, \mathrm{d}y \tag{1}$$

where $f(y)$ is the underlying probability density function of $Y$. Note the difference between diversity and uncertainty: whereas uncertainty describes the state of random variables with incomplete or unknown information, diversity is an attribute of an already known dataset. In general, we can group the data into clusters by any appropriate criterion and estimate $h(Y)$ for every cluster from the $Y$ values in the dataset, thus quantifying the diversity of $Y$ in every cluster. Based on the observation from Figure 1a–b, the OQMD-8 dataset has coverage of materials with diverse $\Delta E$ values in the high symmetry crystal systems, while the $\Delta E$ values in the triclinic and monoclinic systems are not diverse. The J-CFID dataset, on the other hand, lacks diversity in the high-symmetry crystal systems.

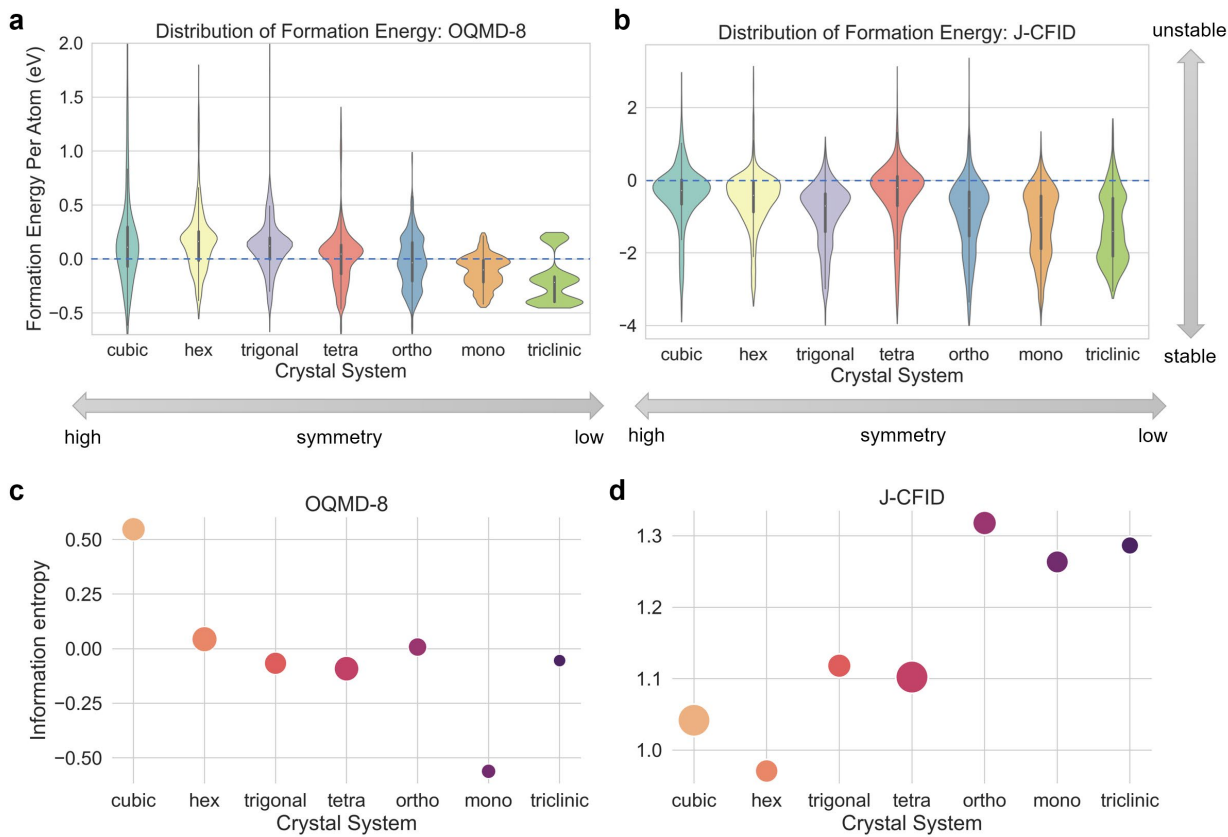

**Figure 1 Structure–stability bias in two datasets. a–b,** Kernel density estimation of the distribution of formation energy among different crystal systems in the OQMD-8 and J-CFID datasets. Without causing ambiguity, only the first few letters are shown for each crystal system for conciseness. **c–d,** Information entropy of formation energy among different crystal systems in OQMD-8 and J-CFID datasets. Colors indicate the degrees of symmetry of crystal systems; the sizes of points reflect the number of datapoints in each crystal system. Note that the OQMD-8 dataset contains only 5 triclinic materials, which causes inaccuracy in the information entropy estimation for the triclinic system.

Next, we measure bias using a fairness criterion[46], i.e., the difference in $h(Y)$ between different clusters indicate the existence and level of bias. For our application, we will use crystal systems as natural clusters, and quantify the structure–stability bias via fairness of $h(\Delta E)$. Figure 1c–d shows that $h(\Delta E)$ captures the observed difference in diversities, thus reflecting the structure–stability bias. The comparison also shows that J-CFID is overall more diverse than OQMD-8, which is because it covers a much larger chemical space. But to measure biases of datasets, the difference of $h(\Delta E)$ between different crystal systems within each dataset is the focus.

## Active Learning for Bias Mitigation

With fairness in diversity as a measure, the data bias can be reduced systematically by adding data to the least diverse crystal system in a manner that increases its diversity in $\Delta E$. We develop the entropy-targeted active learning (ET-AL) algorithm (Figure 2) to attain this automatically. In the active learning context, we refer to the materials with properties known and unknown as "labeled" and "unlabeled", respectively. The ET-AL algorithm iteratively picks a target crystal system (usually the least diverse one), selects an optimal unlabeled material that may improve $h(\Delta E)$ of the system and adds it to the labeled data. The iteration terminates when a pre-specified criterion is satisfied, or all materials are labeled. Details of the algorithm and its implementation are provided in Methods and Algorithm S1.

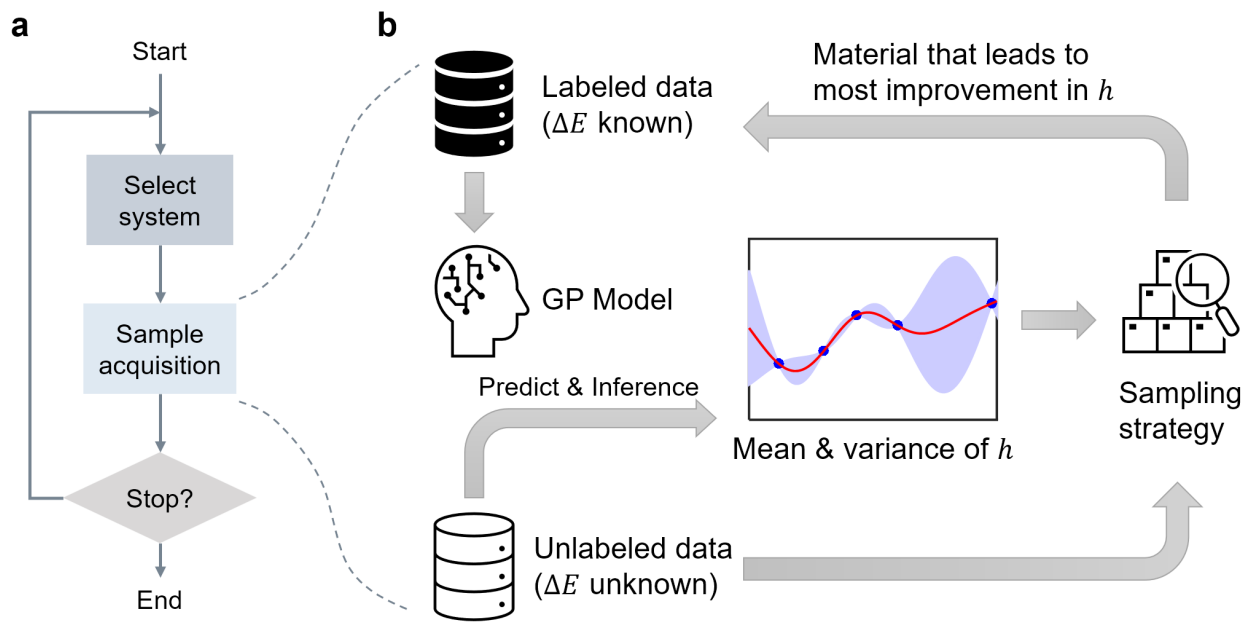

**Figure 2 Schematic of the ET-AL algorithm for data bias mitigation. a**, Overall procedure of ET-AL: a target crystal system is selected, then an unlabeled material is selected and labeled. The steps repeat until the stopping criteria are satisfied. **b**, The procedure of sample acquisition: a Gaussian Process (GP) model is trained with the labeled data and makes predictions for the unlabeled data. The predictive mean and variance of $h$ resulting from adding each material are inferred therefrom. Based on these, the optimal material is selected according to the sampling strategy and added to the labeled data.

## Experimentation and Demonstration

As a demonstration of the ET-AL method, we conduct experiments on the J-CFID dataset. The overall procedure is illustrated in Figure 3a: we split the dataset into a test set, a labeled set with artificial bias, and an unlabeled set. We use ET-AL to augment the labeled set into a low-bias training set (marked `ETAL`) and create another training set (marked `RAND`) of the same size by randomly sampling from the unlabeled set. In addition to demonstrating that ET-AL effectively reduces the structure–stability bias, we show the impact such bias has by comparing supervised ML models for bulk modulus $B$ and shear modulus $G$ derived from the two training sets.

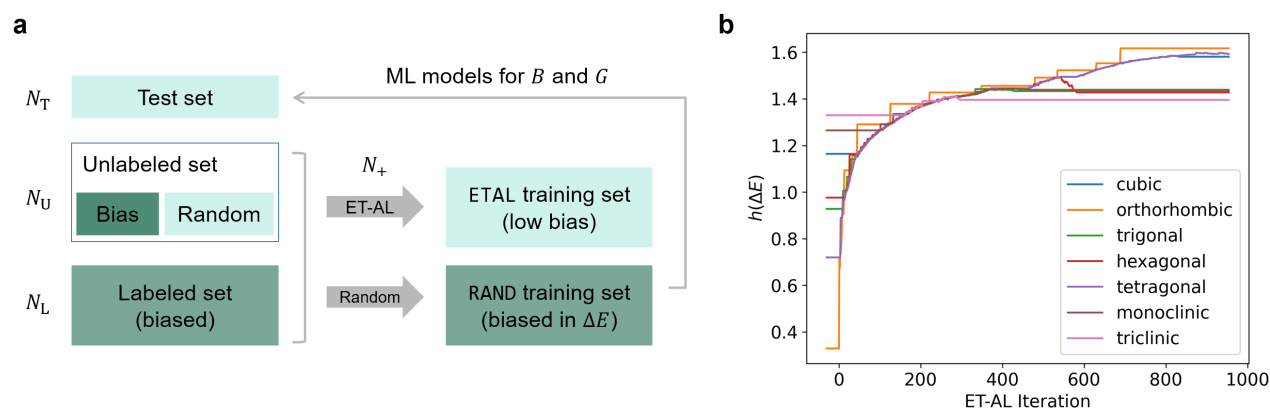

**Figure 3 Experiments on the J-CFID dataset. a,** Split of the dataset: $N_T$ entries are left out as the test set. From the remaining data, some entries are taken away to create an artificial bias and put into the unlabeled set together with randomly selected entries, in total $N_U$. The $N_L$ entries remaining form a labeled set with significant bias. Two training sets are constructed by adding the same number of samples ($N_+$) from the unlabeled set to the labeled set, guided by ET-AL and randomly, respectively. **b,** Change of information entropy in every crystal system during ET-AL iterations. The initial information entropies are shown before iteration 0.

In the experiment, we set $N_L = 1,000$, $N_U = 4,898$, and $N_T = 5,000$. The artificial bias is introduced by removing all tetragonal and trigonal materials with $\Delta E > 0$ and all orthorhombic materials with $\Delta E < 0$. With materials represented by graph embeddings (presented in Methods), ET-AL is applied to the dataset and runs for 985 iterations before termination. As Figure 3b shows, the introduced bias is captured by the diversity metric (the three manipulated crystal systems have relatively low initial $h$), and mitigated by ET-AL. Moreover, through ET-AL, the dataset reaches a state where diversities of crystal systems are closer to each other as compared to the initial state, which is favored by the fairness criterion (also shown in Figure S1). A similar demonstration performed on the OQMD-8 dataset is presented in Figure S2 and Figure S3.

Next, we investigate the effects of ET-AL on dataset distribution. We employ t-distributed stochastic neighbor embedding (t-SNE)[47] for dimension reduction of the graph embedding representations of J-CFID data into a 2-dimensional space. The low-dimensional embeddings acquired by t-SNE reflect the distribution of data in the structure space. In Figure 4a, we use these embeddings to show the coverage of the labeled dataset (see also Figure S4) and the ET-AL-selected and randomly selected data (b–c). ET-AL guides sampling in the underrepresented regions (lighter shades in Figure 4a), as opposed to a nearly uniform coverage by random sampling in Figure 4c.

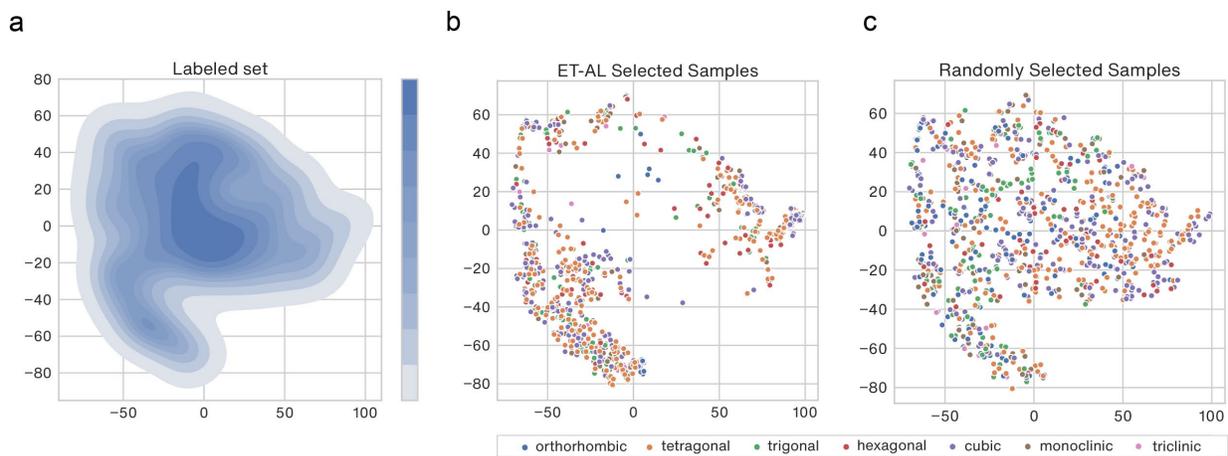

**Figure 4 Visualization of dataset distributions. a**, Kernel density estimation (KDE) plot of tSNE embeddings of the labeled dataset. The shade shows the density of points, regions with lighter colors are less covered. **b–c**, t-SNE plots of graph embeddings of the materials selected by ET-AL and random sampling, respectively, with colors indicating crystal systems. Compared to random sampling, materials selected by ET-AL better cover the region where labeled samples are sparse, as well as the crystal systems where artificial bias is introduced.

To assess the impact of bias on property prediction, we train multiple supervised learning models on the two training sets, both of size 1,954, for predicting $B$ and $G$ from a set of physical descriptors (detailed in Methods). Each model is trained 30 times with different random states (controlling the initialization, feature permutation, etc., but not affecting training data), and the coefficient of determination ($R^2$) on the test set is recorded. Models include random forest (RF), gradient boosting (GB), neural network (NN), and support vector regression (SVR), among which RF and GB attain relatively better performances on the task. A potential reason for such performance difference is that the descriptors form heterogeneous tabular data, for which tree ensemble models have an advantage[48]. We summarize the performances of these ML models in Figure 5, from which we find that models derived from the ETAL dataset with reduced bias display systematically superior accuracies over those from the RAND dataset.

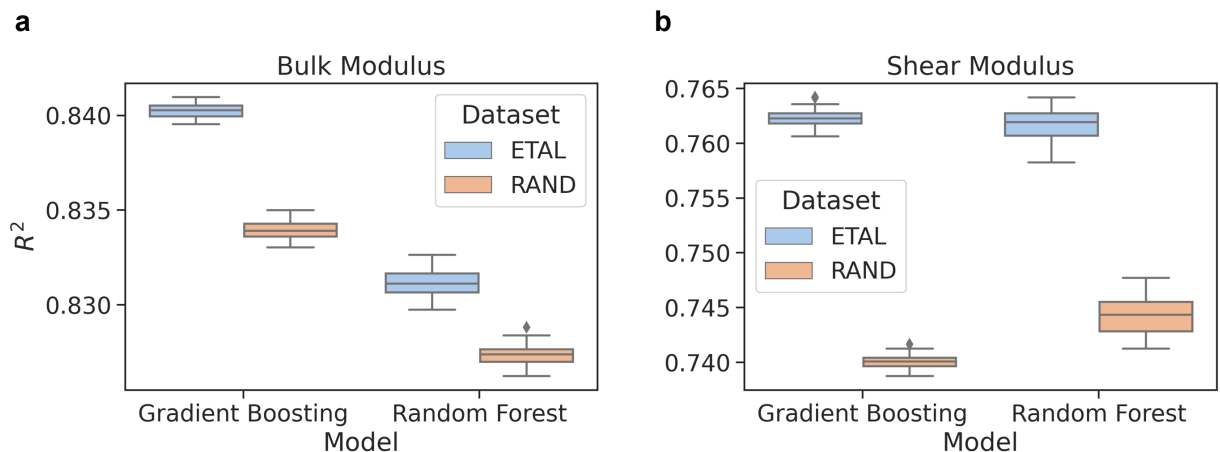

**Figure 5 Comparison of supervised ML models trained on two J-CFID-derived datasets.** The boxplots show testing $R^2$ of each model type and target property across 30 replicates, **a**, for bulk modulus and **b**, for shear modulus. Models trained on the ETAL dataset display higher testing $R^2$ than those trained on the RAND dataset.

In Figure 6, we mark the "most improved samples", i.e., testing samples for which the ML models' prediction accuracies using the ETAL training set show greater advantages compared to using the RAND training set. As observed, most of these samples are in the underrepresented regions of the labeled set (low-density regions in Figure 4a). ET-AL's focus in these regions during sample selection (triangles in Figure 6 overlap with sampling points in Figure 4b) leads to the better accuracy of ML models trained on the `ETAL` dataset. These observations agree with the findings of Li et al.[49]: ML models trained on a biased dataset lack generalizability to underrepresented test samples. ET-AL provides a solution to the problem by reducing structure–stability bias, which improves the coverage of the dataset in the structure space, and thus facilitates downstream tasks such as ML modeling of mechanical properties $B$ and $G$.

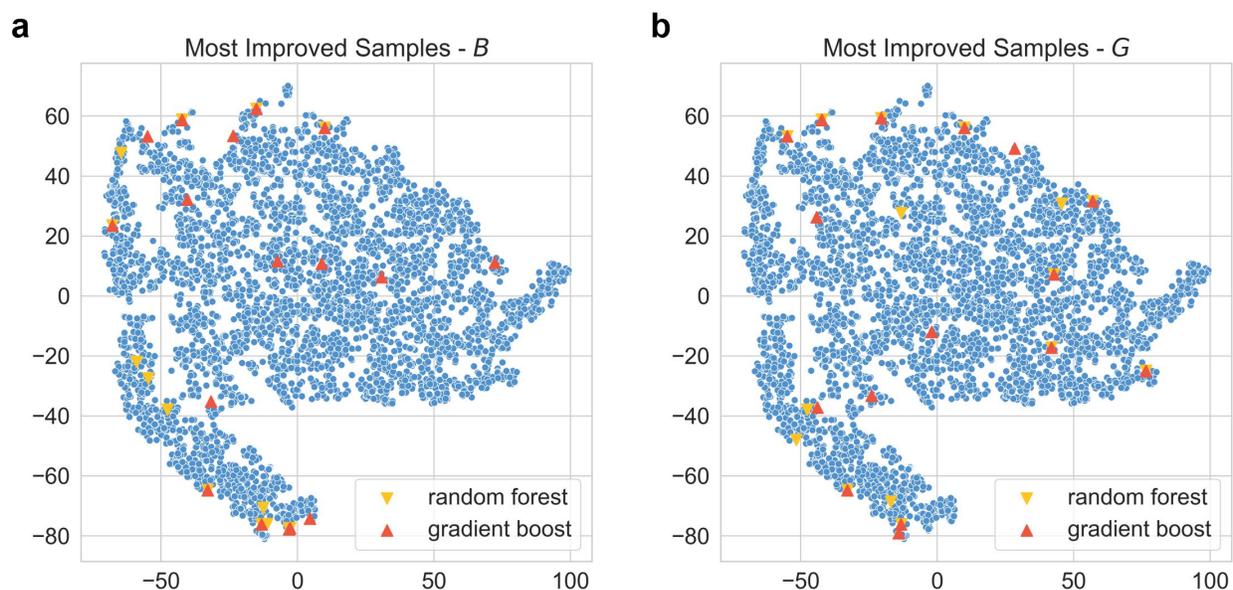

**Figure 6 Locations of the most improved samples.** Blue dots show the tSNE mapping of the test samples' graph embeddings. The triangles mark the samples where ML models trained on the `ETAL` dataset show the most advantages in accuracy over those trained on the `RAND` set, for **a**, bulk modulus and **b**, shear modulus. 20 samples are shown for each model–property combination.

## Potential Applications

The data bias metric and ET-AL method proposed in this work have a wide range of applications in materials discovery and beyond. First, researchers may examine and potentially reduce the bias in their datasets before developing data-driven models thereon or publishing the data. Second, ET-AL allows steering autonomous data acquisition in an unbiased way. This includes high-throughput computation, as well as experiments such as self-driving laboratories[50]. Though we presented the structure–stability bias as an example, the method applies to other forms of bias as well.

An application of particular significance is dealing with bias in materials data resources. Since new materials are continually added to the databases, ET-AL can fit in the pipeline to select the materials to add. In practice, however, some databases are so large that an observable effect of bias mitigation requires adding many new data points, and there are other considerations besides bias in database construction. In remediation, the information entropy-based bias metric can also guide trimming rather than expanding a database, i.e., selecting a less biased subset. The level of bias can be tuned according to the need of usage.

Though originally proposed for materials data, ET-AL is generally applicable to other fields where large data are generated and curated for future reuse[51]. Protein database[52] for biomedical studies, and geometry datasets[53] for design and manufacturing studies are among a few examples. Bias in these data may lead to inaccuracies in parameter calibration, predictive modeling, or design optimization, and ET-AL enables detection and mitigation of the bias.

## CONCLUSION

We highlighted the previously overlooked bias in materials data resources, which has an impact on a broad range of data-driven materials modeling and design studies. We proposed a generic metric for data bias based on diversity measured by information entropy, which successfully captures the structure–stability bias in datasets retrieved from widely used materials data platforms OQMD and JARVIS. We then formulated and implemented an entropy-target active learning (ET-AL) framework to automatically reduce bias in datasets by acquiring new samples. Through ablation studies, we demonstrated that ET-AL can effectively reduce the structure–stability bias, thus improving data coverage in the structure space and increasing the accuracy of data-driven modeling of materials properties.

We also note that as a generic framework, ET-AL's capability is not limited to materials databases. As the data-driven research paradigm has been adopted by various domains, and data bias is ubiquitous in almost every data system, we anticipate that the ET-AL method is applicable to a variety of scientific and engineering domains, to facilitate the curation of high-quality data and data-driven studies.

## METHODS

### Dataset Preparation

**Data collection and cleaning.** The OQMD-8 dataset is retrieved from OQMD using its API implemented in the "qmpy-rester" package. The J-CFID dataset is downloaded from figshare.com[42]. Out of the >50,000 entries, the ones reporting positive $B$ and $G$ values are kept. Entries containing elements H, Tc, halogens (VIIA), noble gases (VIIIA), lanthanum family, and those with atomic numbers ≥84 are excluded. Figure S5 and Figure S6 show some statistics of the J-CFID dataset.

**Graph embedding representation.** We use the crystal graph convolutional neural network (CGCNN) model[54], which maps a graph encoding of crystal structure to a numerical vector before predicting its properties. The numerical vector provides a representation of materials' structures regulated by the target property. We feed the crystallographic information framework (CIF) files of all entries in the dataset to a pretrained CGCNN model and extract a 32-dimensional vector representation for each material's structure. Unless specified otherwise, we use the model pretrained on formation energy per atom. The graph embeddings of materials are used as inputs for model training and evaluation of unlabeled materials in ET-AL.

Besides graph embeddings, many other representations that can be derived from materials' crystal structures without knowing their properties are also compatible with ET-AL, examples include fragment descriptors[55] and tensor representations[56].

## Information Entropy

Information entropy of continuous-valued $\Delta E$ is estimated from a discrete set of $\Delta E$ values using the "differential_entropy" function from the scipy.stats package[57]. The software automatically selects a numerical method for entropy estimation[58] based on data size; details of the numerical estimation methods are described in Supplementary Materials.

## Active Learning

**Target system selection.** In every iteration, all crystal systems with unlabeled sample(s) available are candidates. Systems that are sampled but not improved five consecutive times are excluded. Of the remaining candidates, the crystal system with the lowest $h(\Delta E)$ is selected as the target.

**Gaussian process modeling.** Gaussian Process (GP) modeling[59] builds upon the assumption that the responses $\boldsymbol{y} = \{y_1, \dots, y_m\}$ are jointly Gaussian distributed given the predictors $\boldsymbol{X} = \{\boldsymbol{x}_1, \dots, \boldsymbol{x}_m\}$:

$$\boldsymbol{y}|\boldsymbol{X} \sim \mathcal{N}(\boldsymbol{\mu}, \boldsymbol{K} + \sigma^2 \boldsymbol{I}) \tag{2}$$

where the covariance matrix $\boldsymbol{K}$ is inferred from the similarity between predictors using a kernel function, and the $\sigma^2 \boldsymbol{I}$ term accounts for noise. Once trained, inputting an unseen predictor $\hat{\boldsymbol{x}}$, the model outputs not a single value but a predicted Gaussian distribution of the response $\hat{y}$. Hence, GP is an uncertainty-aware machine learning model. We train GP models using the "ExactGPModel" module of the GPyTorch package[60], with the graph embedding representation as predictors and $\Delta E$ as response.

**Monte Carlo inference.** For each unlabeled material, the GP model provides a predicted distribution of $\Delta E$, from which we use the Monte Carlo method[61] to infer the resulting change in $h(\Delta E)$ by adding the material, as illustrated in Figure 7. We thereby obtain the predictive mean and variance of $h$ for every unlabeled material, which are later used in the evaluation of the sampling criterion.

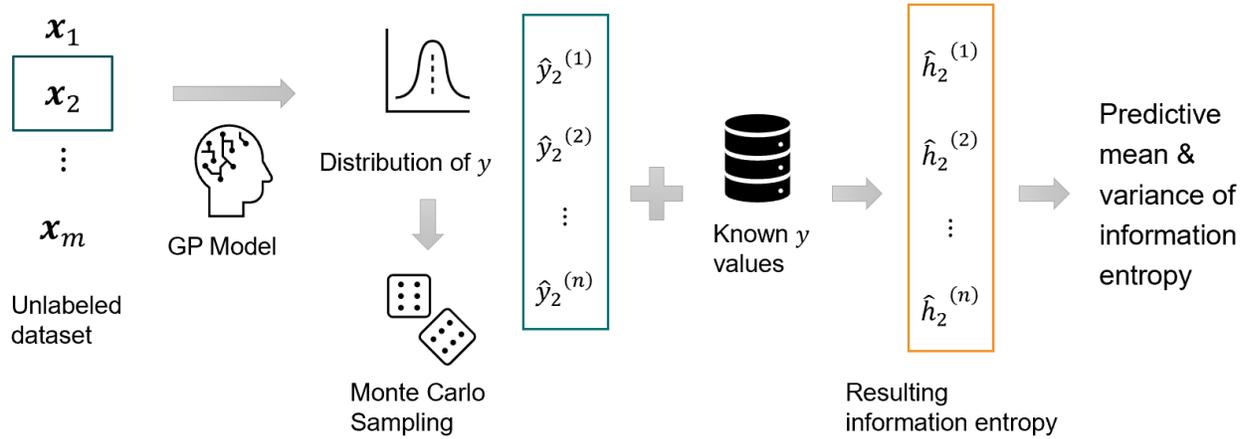

**Figure 7 Schematic of the Monte Carlo Inference for mean and variance of $h$.** For simplicity, formation energy or any property of interest is denoted as $y$ in the figure. For each unlabeled material, we obtain a predicted distribution of $y$ via the GP model, and randomly draw $n$ samples therefrom. By trying to add every sample into the $y$ values of labeled dataset, we get $n$ resulting $h$ values ($n = 1,000$ is used in experiments), from which we can obtain the predictive mean and variance of the resulting $h$ if that material is selected.

**Sampling strategy.** The selection of unlabeled materials is guided by the expected improvement (EI) sampling criterion[62]. For every unlabeled material $\boldsymbol{x}$, the expected improvement in $h$

$$\text{EI}(\boldsymbol{x}) = \text{E}[\max\{0, \Delta(\boldsymbol{x})\}] = \hat{s}(\boldsymbol{x})\phi\left(\frac{\Delta(\boldsymbol{x})}{\hat{s}(\boldsymbol{x})}\right) + \Delta(\boldsymbol{x})\Phi\left(\frac{\Delta(\boldsymbol{x})}{\hat{s}(\boldsymbol{x})}\right) \qquad (3)$$

is evaluated, where $\Delta(\boldsymbol{x}) = h(\boldsymbol{x}) - h_{\text{cur}}$ is the difference between the predicted mean $h$ and the current $h$; $\hat{s}(\boldsymbol{x})$ is the predicted standard deviation of $h$; $\phi(\cdot)$ and $\Phi(\cdot)$ are the probability density function (*pdf*) and cumulative distribution function (*cdf*) of standard Gaussian distribution, respectively. The unlabeled material with the largest EI is selected, i.e., $\boldsymbol{x}^* = \arg\max_{\boldsymbol{x}} \text{EI}(\boldsymbol{x})$.

More generally, the selection operates in batches, i.e., one or multiple unlabeled material(s) with large EI are selected in every iteration. A batch size of 1 is used in the implementation of this work, as the time for running an ET-AL iteration is negligible compared to the acquisition of a new datapoint (e.g., first-principles calculation, experimental measurement). On the other hand, data acquisition can be parallelized. In that case, ET-AL can be easily configured to use a larger batch size, thus further improving computational efficiency.

**Stopping criteria.** In our experiments on the J-CFID dataset, the active learning process is terminated when the target crystal system has the highest $h(\Delta E)$ of all systems. In that case, improving its $h$ will worsen the fairness of diversity. However, when evaluation (e.g., first-principles calculation) of new materials is feasible, the existence of materials improving $h$ of the least diverse system is almost guaranteed. In such application scenarios, stopping criteria may be specified according to resources and budget, or may not be needed.

## Supervised Machine Learning

**Materials featurization.** Compositional and structural descriptors are retrieved for every crystal structure using an automatic workflow developed by Georgescu et al.[63], built upon the pymatgen[64] and Matminer[65] toolkits. Descriptors that contain NaN values or have zero variance among the dataset are removed, resulting in 117 descriptors for every material, which are then used as inputs for supervised ML models.

**Machine learning.** The supervised ML models are trained and tested using implementations in the scikit-learn package[66]. The hyperparameters are selected using a 5-fold cross-validated (CV) grid search; unless otherwise specified, they are the same as the default settings of the package. For SVM, the radial basis function (RBF) kernel is used. For neural networks, model architecture and hyperparameters are tuned: 1 hidden layer with 128 neurons, $L_2$ regularization strength $\alpha =$ 0.1, rectified linear unit (ReLU) activation for hidden layers, and Adam optimizer. For random forests (RF), the number of trees is set to 300, and the maximum depth of individual trees is set to 30 for predicting $G$ and 100 for predicting $B$. For gradient boosting (GB), the maximum depth of individual trees is 5. Detailed hyperparameters of the RF and GB models are listed in Table S1.

## ACKNOWLEDGEMENTS

This work was supported by the Advanced Research Projects Agency-Energy (ARPA-E), U.S. Department of Energy, under Award Number DE-AR0001209, and the Center for Hierarchical Materials Design, under Award Number ChiMaD NIST 70NANB19H005. The views and opinions of authors expressed herein do not necessarily state or reflect those of the United States Government or any agency thereof. We also acknowledge Francesca M. Tavazza and Brian L. DeCost for assistance with data collection. H.Z. acknowledges Kyle Miller, Dale Gaines II, Adetoye H. Adekoya, Whitney Tso, and G. Jeffrey Snyder for insightful discussions, and Ke Sun for valuable advice on visualization.

## AUTHOR DECLARATIONS

### Author Contributions

**H.Z.:** Conceptualization, Formal Analysis; Investigation; Methodology; Software; Visualization; Writing – Original Draft; Writing – Review & Editing. **W.W.C.:** Formal Analysis; Methodology; Software; Writing – Review & Editing. **J.M.R.** and **W.C.:** Formal Analysis; Writing – Review & Editing; Supervision; Funding Acquisition.

### Conflict of Interest

The authors have no conflicts of interest to disclose.

## DATA AVAILABILITY

The datasets used in bias characterization and experimentation are openly available at oqmd.org (OQMD-8 dataset) and reference number[42] (J-CFID dataset). The code and data that support the findings are openly available on the ET-AL GitHub repository at https://github.com/Henrium/ET-AL and Zenodo at https://doi.org/10.5281/zenodo.7406332 .

# Supplementary Items

Algorithm S1 Pseudocode of the ET-AL algorithm.

---

**Entropy-targeted active learning.**

---

**Input:** Labeled dataset $\mathcal{D} = \{\boldsymbol{x}_i, y_i\}$, unlabeled dataset $\mathcal{U} = \{\boldsymbol{x}_j'\}$; $\mathcal{D} = \bigcup_c \mathcal{D}_c$, $\mathcal{U} = \bigcup_c \mathcal{U}_c$ ( $c \in$ crystal systems); Monte Carlo sample size $n$; stopping criteria

   **while** stopping criteria are not satisfied **do**

      Calculate information entropies $H(\mathcal{D}_c) = h(y \in \mathcal{D}_c)$, select $c^* = \arg\min_c H(\mathcal{D}_c)$

      Fit a GP to $\mathcal{D}_{c*}$: $Y \sim GP(\boldsymbol{X})$

      **for** $\boldsymbol{x}_j' \in \mathcal{U}_{c*}$ **do**

         Draw $n$ samples $\left\{y_j^{(k)}\right\}_{k=1}^n$ from $GP(\boldsymbol{x}_j')$

         Calculate $h_j^{(k)} = H\left(\mathcal{D} \cup \left\{y_j^{(k)}\right\}\right)$ **for** $k = 1$ **to** $n$

         Calculate the mean and variance of $\left\{h_j^{(k)}\right\}_{k=1}^n$, $\mathrm{EI}(\boldsymbol{x}_j')$ according to Equation (3)

      **end for**

      Select sample $\boldsymbol{x}^* = \arg\max_{\boldsymbol{x}' \in \mathcal{U}_{c*}} \mathrm{EI}(\boldsymbol{x}')$

      Acquire $y^*$, remove $\boldsymbol{x}^*$ from $\mathcal{U}_{c*}$, add $\boldsymbol{x}^*, y^*$ to $\mathcal{D}_{c*}$

   **end while**

---

**Returns:** Updated dataset $\mathcal{D}$

---

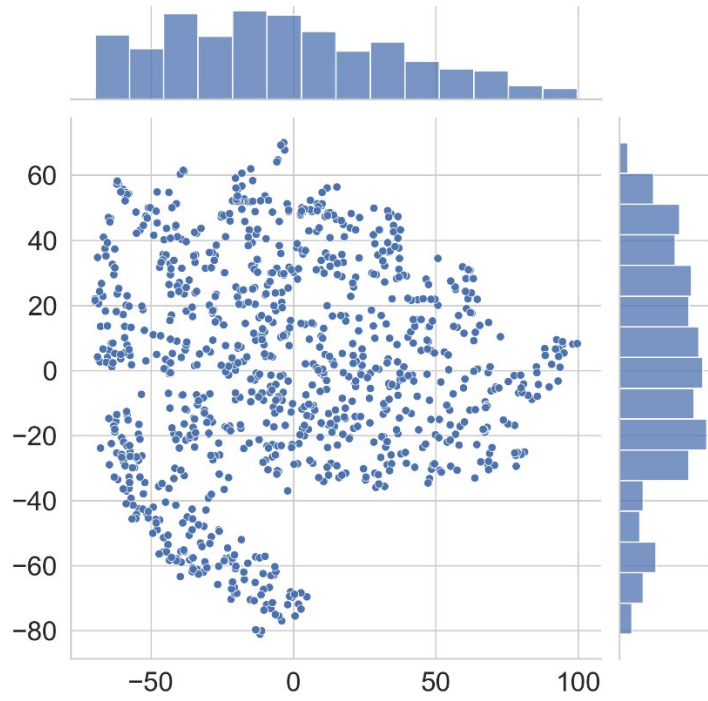

Figure S1 t-SNE plot of graph embeddings of the labeled dataset. The histograms show distribution densities in different regions.

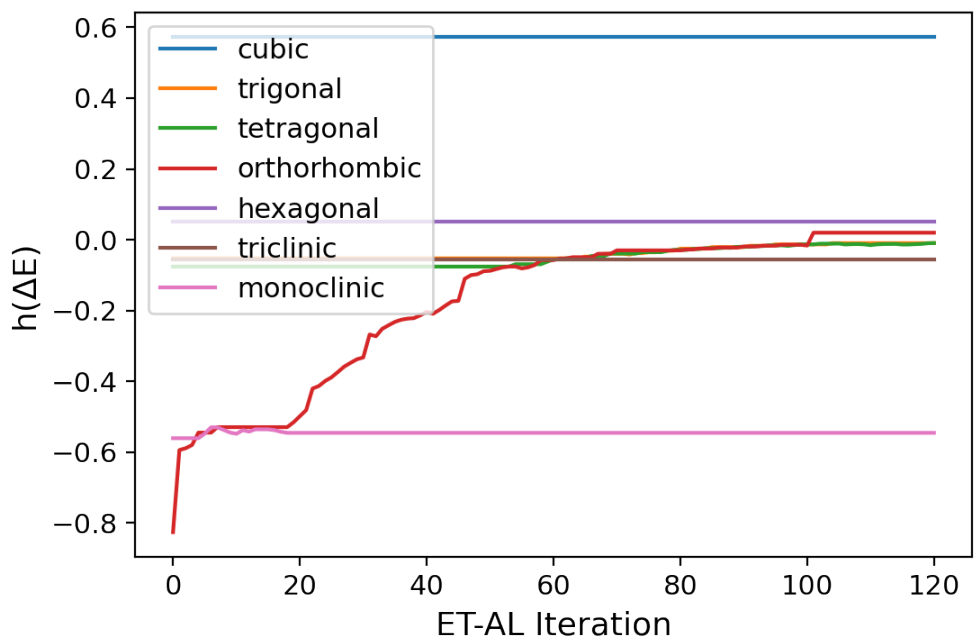

Figure S2 Change of information entropy in every crystal system during ET-AL iteration on the OQMD-8 dataset. Artificial bias is introduced by leaving out all orthorhombic compounds with $\Delta E > 0$, together with randomly selected compounds, in total 1,000, as the unlabeled dataset. $h(\Delta E)$ for monoclinic is not improved in later iterations because new materials are selected only in the unlabeled dataset, which is limited by the diversity of OQMD-8 data.

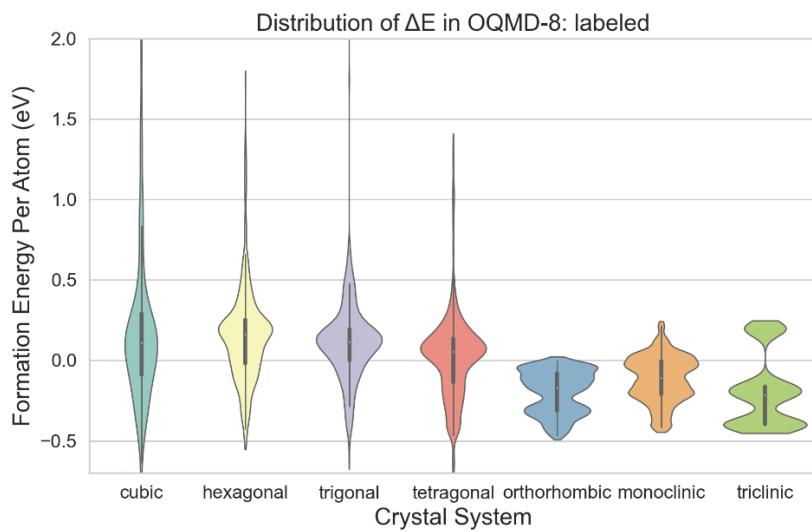

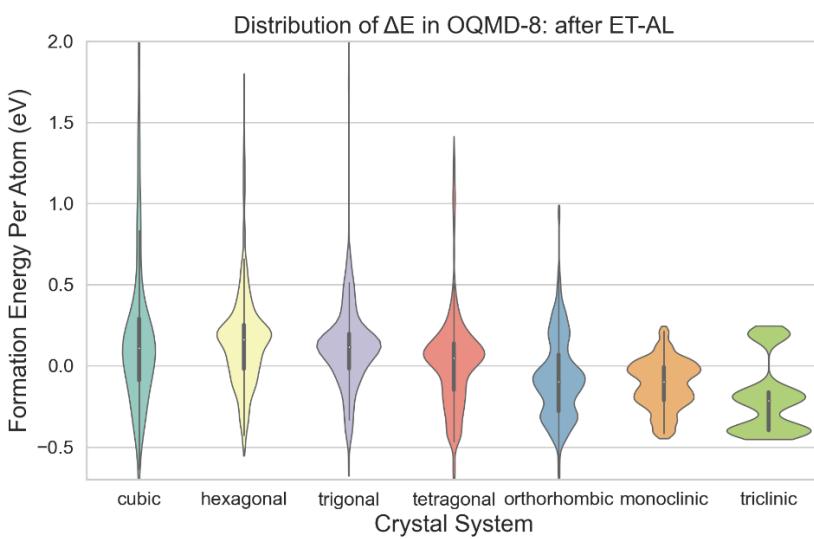

Figure S3 Kernel density estimation of the distribution of $\Delta E$ among different crystal systems in the OQMD-8 derived datasets: (top) before and (bottom) after ET-AL.

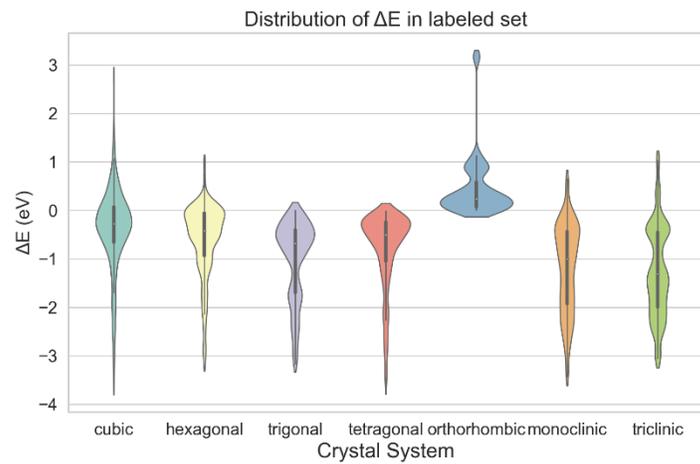

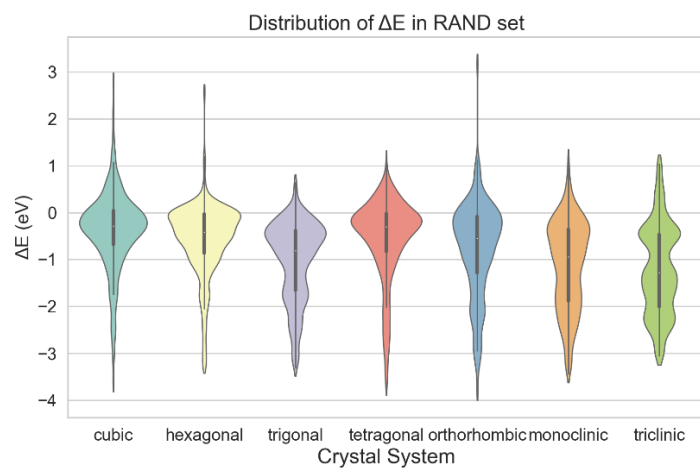

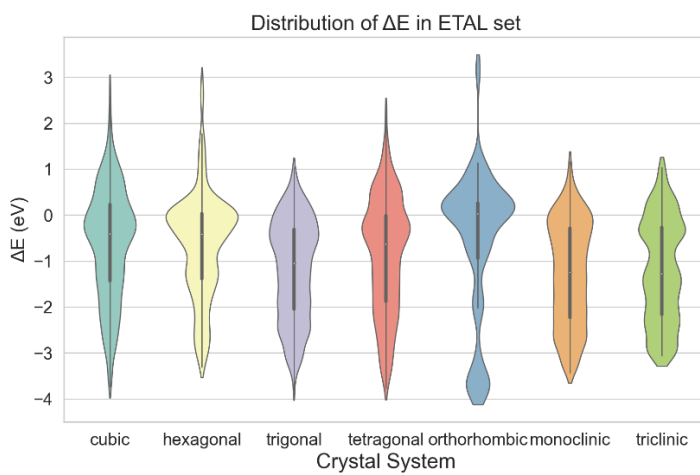

Figure S4 Kernel density estimation of the distribution of $\Delta E$ among different crystal systems in the J-CFID-derived (from top to bottom) labeled, RAND, and ETAL datasets.

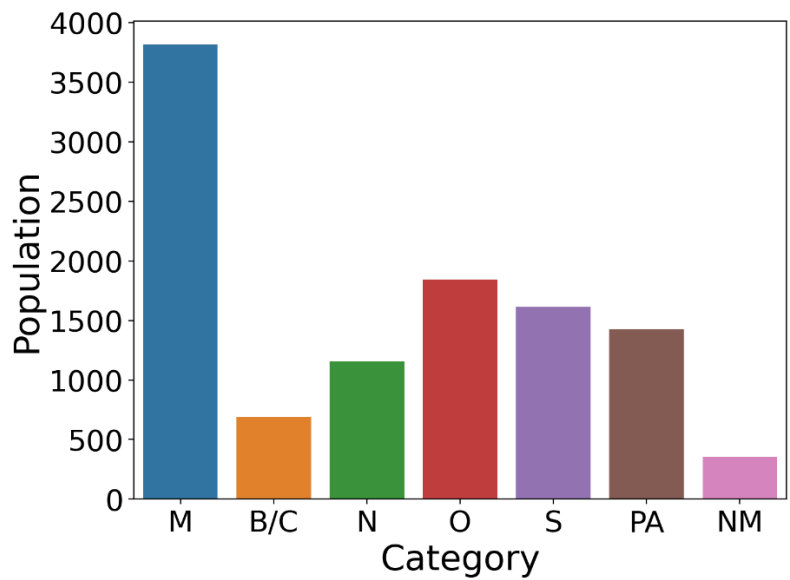

(a) The population of the J-CFID dataset by categories. M: metal or intermetallic. Columns 2–5 are compounds with metallic element(s) and a single type of anion: B/C: B/C/Si anion; N: N/P/As anion; O: O anion; S: S/Se/Te anion. PA: polyanionic, i.e., compounds with more than one type of anions above. NM: nonmetal, i.e., compounds without metallic elements.

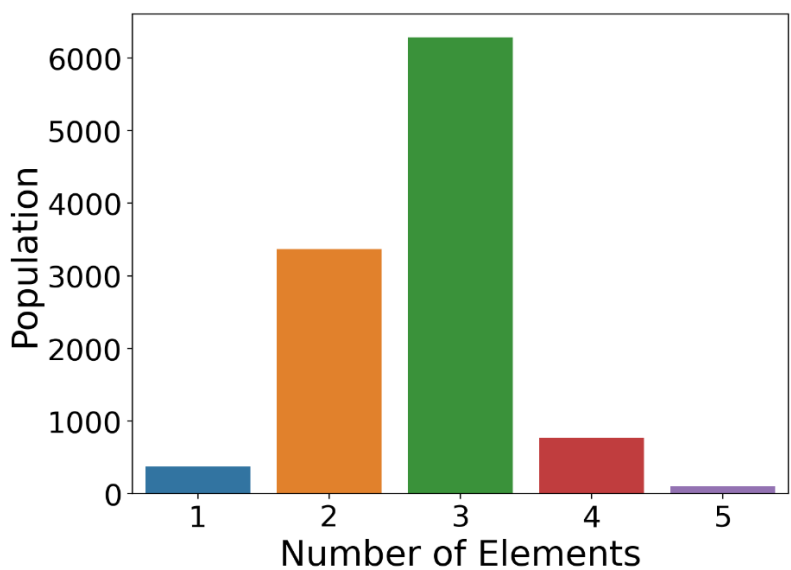

(b) The population of the J-CFID dataset by the number of component elements.

Figure S5 Statistics of the J-CFID dataset by number and types of component elements.

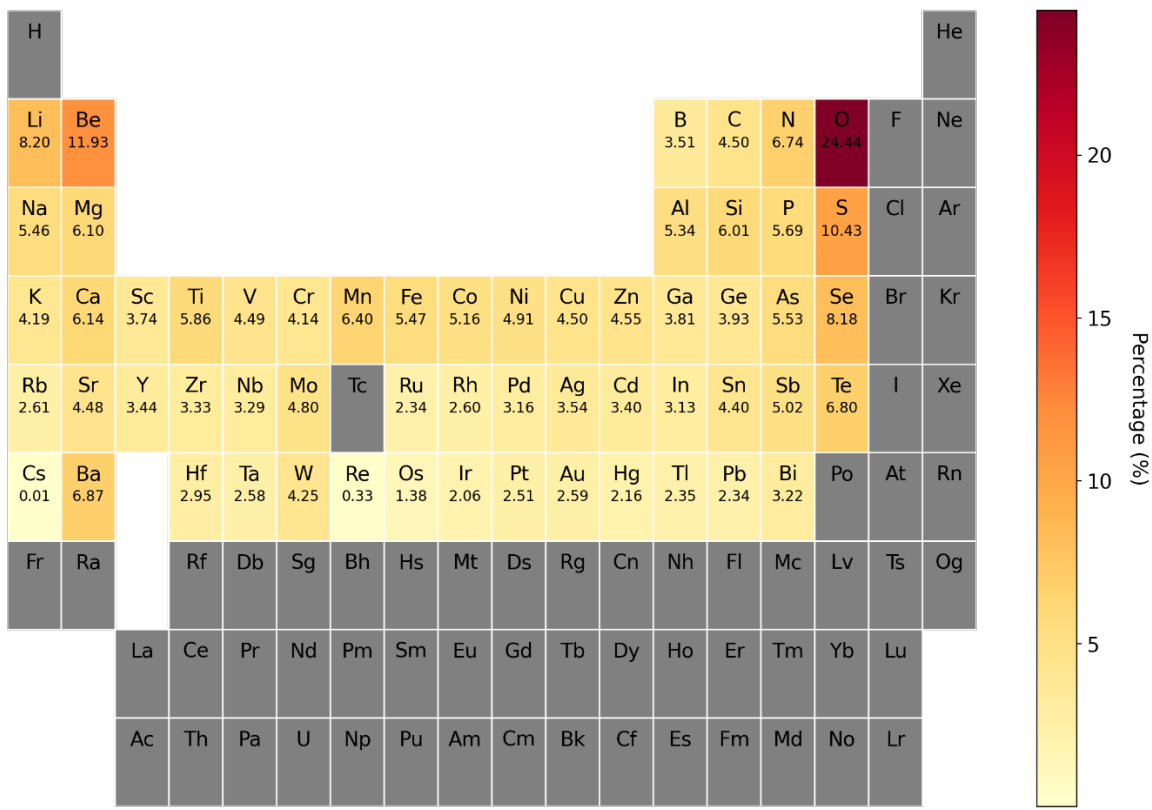

Figure S6 Percentage of compounds containing each element in the J-CFID dataset.

Table S1. Tuned hyperparameters of random forest and gradient boosting models for predicting $B$ and $G$. "^" denotes that the effect of a hyperparameter is found to be insignificant, hence, the default setting is used.

| Hyperparameters | Random Forest | | Gradient Boosting | |
|---|---|---|---|---|
| | $B$ | $G$ | $B$ | $G$ |
| bootstrap | True | True | True | True |
| max_depth | 100 | 30 | 5 | 5 |
| max_features | n_features | n_features | n_features | n_features |
| min_samples_leaf | 1^ | 1^ | 1^ | 1^ |
| min_samples_split | 2^ | 2^ | 2^ | 2^ |
| n_estimators | 300 | 300 | 100 | 100 |

# Supplementary Text

## 1. Bias metric

### Metric selection

One definition of data bias is "unjustifiable concentration on a particular part"[1]. In the context of this work, the "parts" can be regions in a space broadly defined by composition, (micro)structure, property, processing, or energy-based descriptions. As an example, the structure–stability bias arises from concentration (uneven coverage) in materials with certain structures (crystal systems) and stability ($\Delta E$). Such uneven coverage can be captured by the different diversities of stability among different crystal systems. The information entropy

$$h(Y) = -\int f(y) \log f(y) \mathrm{d}y$$

has been widely adopted as a metric for diversity[2]. Its numerical nature and simplicity in the calculation are desirable as a target in active learning. Also widely adopted as a numerical diversity measure is the determinantal point processes (DPP)[3]. However, DPP is evaluated pairwise, thus lacking scalability to large materials databases.

Another seemingly applicable metric is the conditional information entropy

$$h(Y|CS) = -\int f(cs, y) \log f(y|cs) \mathrm{d}cs \mathrm{d}y$$

where $CS$ denotes the crystal system. However, this metric has several problems. First, $f(cs, y) = f(y|cs) \cdot f(cs)$, where $f(cs)$, the probability distribution of $CS$, is related to the fractions of crystal systems. The "true" fractions in nature are unknown; besides, in the examination of bias, we focus on the coverage evenness rather than the populations of crystal systems. Second, $h(Y|CS) \leq h(Y)$, and equality is reached (i.e., active learning is concluded) if and only if $Y$ and $CS$ are independent. Intuitively, with $Y$ denoting $\Delta E$, this means symmetry does not provide information for stability. This is not true according to theories discussed in the introduction of structure–stability bias. Third, $h(Y|CS)$ as a single scalar value does not give a sense of the bias level, whereas the difference in $h(Y)$ of crystal systems indicates how biased a dataset is.

### Information entropy estimation

For a continuous random variable $y$, evaluation of its information entropy requires the probability density function ($pdf$) $f(y)$. However, given a discrete set of values $\{y_i\}_{i=1}^{n}$, the underlying $pdf$ is not obtainable. We use the numerical estimations implemented in scipy.stats, with the "auto" setting. Specifically, for $11 <$ $n \leq 1000$, the result is given by the $H_e(m, n)$ estimator presented by Ebrahimi et al.[4]; while for $n > 1000$, the Vasicek estimator[5]

$$h(Y) \cong \frac{1}{n} \sum_{i=1}^{n} \log\left(\frac{n}{2m}(y_{i+m} - y_{i-m})\right)$$

is used, where $m$, the window size, is defined by $\lfloor \sqrt{n} + 0.5 \rfloor$ in both cases.

## 2. Data preparation

### Graph embedding

The crystal graph convolutional neural network (CGCNN)[6] predicts materials properties from the graph representation of the crystal structures. The input includes node feature vector(s) that encode atomic properties, and edge feature vector(s) that encode connections between atoms, both can be obtained from crystallographic information framework (CIF) files without knowing other properties.

We retrieve a CGCNN model pretrained to predict the formation energy per atom ($\Delta E$) on the Materials Project dataset[7]. We feed the CIF files of materials in the J-CFID to the pretrained model and obtain the activations of the last but one layer of neurons. These 32-dimensional vectors (graph embeddings) are used as representations of J-CFID materials structures. Note that graph embeddings do not have direct physical meanings, but they are generally obtainable for any given crystal structure.

### Physical descriptors

In supervised machine learning (ML) of mechanical properties, we use physically meaningful descriptors of materials as input. The descriptors include ones defined in the Magpie ML framework[8], the Ewald energy per atom, and volume per site. The Magpie descriptors set include the minimum, maximum, range, mean, average deviation, and mode of features such as Mendeleev number, atomic weight, and covalent radius of elements/atoms in a compound. A complete list can be found in the data files open-sourced along with the code.

The Ewald energy per atom is obtained using the "analysis.ewald" module of the pymatgen package[9]; the other descriptors are obtained using the "featurizers" module of the Matminer package[10].

### J-CFID data splitting

In preparation of the data for experiments, we first split the J-CFID dataset of size 10,898 to a test set of 4,898 datapoints and an "experiment" set of 6,000 datapoints. Due to the large size and randomness in dataset splitting, the test set has a similar level of bias as the original J-CFID dataset. From the experiment set, we take (1) all tetragonal and trigonal materials with $\Delta E > 0$ and (2) all orthorhombic materials with $\Delta E < 0$, then randomly draw datapoints from the remaining data, to form an unlabeled set of size 5,000. The other 1,000 datapoints make the labeled set.

## 3. Algorithm Analysis

The execution of ET-AL consists of three main parts: (1) Gaussian Process (GP) model fitting on the labeled data, (2) selection of an unlabeled datapoint, and (3) label acquisition. The time part (3) takes depends on the experimental/computational technique used, and it is usually the most time-consuming component, compared to which the time of (1) and (2) are negligible. Quantities affecting the computational time of parts (1) and (2) include the size of the labeled dataset $N_{\mathrm{L}}$, size of the unlabeled dataset $N_{\mathrm{U}}$, and the number of Monte Carlo samples $n_{\mathrm{MC}}$.

For part (1), GP fitting, the time complexity is $\mathcal{O}(N_{\mathrm{L}}^3)$. The scalable GP models[11] provide a solution for better scalability on large datasets. In particular, the sparse variational (SV) GP model[12] reduces the time complexity to $\mathcal{O}(m^3)$ (where $m \ll N_{\mathrm{L}}$ is the number of inducing points whose distribution is representative of the training data), with a slight loss of accuracy.

In part (2), for every unlabeled data, the GP model makes one prediction, costing computational time $\mathcal{O}(N_{\mathrm{U}})$; then $n_{\mathrm{MC}}$ Monte Carlo samples are drawn, and for each sample the information entropy and acquisition function are calculated, costing computational time $\mathcal{O}(N_{\mathrm{U}} \cdot n_{\mathrm{MC}})$. The total time complexity of part (2) is $\mathcal{O}(N_{\mathrm{U}}(k + n_{\mathrm{MC}}))$, where $k$ is constant.

# Supplementary References